\documentclass[
journal=nalefd, 
manuscript=article,layout=twocolumn]{achemso}
\setkeys{acs}{articletitle = true}

\usepackage[colorlinks=true,linkcolor=black,citecolor=black]{hyperref}
\usepackage{color,soul} 
\usepackage[utf8]{inputenc}
\usepackage{booktabs}
\usepackage{epstopdf}
\usepackage{graphicx}
\usepackage{longtable}
\usepackage[version=3]{mhchem} 
\usepackage{multirow}
\usepackage{footnote}
\makesavenoteenv{tabular}
\makesavenoteenv{table}

\author{Weiwei Gao}
\email{weiwei@ices.utexas.edu}
\author{James R. Chelikowsky}
\email{jrc@utexas.edu}
\affiliation[1]{Center for Computational Materials, Oden Institute for Computational Engineering and Sciences, The University of Texas at Austin, Austin, TX 78712}
\alsoaffiliation[2]{Department of Physics, The University of Texas at Austin, Austin, TX 78712}
\alsoaffiliation[3]{McKetta Department of Chemical Engineering, The University of Texas at Austin, Austin, TX 78712}

\title[\texttt{achemso} demonstration]
{Prediction of intrinsic ferroelectricity and large piezoelectricity in monolayer arsenic chalcogenides}

\begin{document}

\begin{tocentry}
	\includegraphics{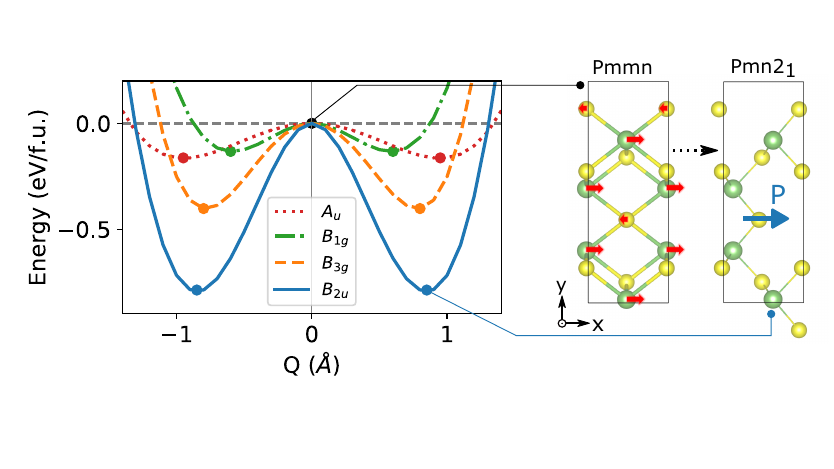}
\end{tocentry}

\begin{abstract}
Two-dimensional materials that exhibit spontaneous electric polarization are of notable interest for functional materials.
However, despite many two-dimensional polar materials are predicted in theory, the number of experimentally confirmed two-dimensional ferroelectrics are still far less than bulk ferroelectrics. 
We provide strong evidence that the Pmn2$_1$ phase of arsenic chalcogenides As$_2$X$_3$ (X=S, Se, and Te), which include the recently isolated monolayer orpiment, are intrinsic ferroelectrics and demonstrate strong in-plane piezoelectricity. 
We found the calculated energy barriers for collectively reversing the electric polarization or moving a 180$^\circ$ domain wall are reasonable compared to previously reported ferroelectrics.
We propose a high-symmetry structure (with Pmmn space group) transforms into the ferroelectric Pmn2$_1$ phase by a soft B$_{2u}$ phonon mode. 
By studying other soft modes of the high-symmetry Pmmn structure, we identify several undiscovered metastable polymorphs, including a polar phase (with a P2$_1$ space group) with sizable piezoelectricity. 
\end{abstract}

\section{Introduction}









Materials lacking inversion symmetry may display useful properties such as piezoelectricity and ferroelectricity, which have wide applications in modern industries. 
In particular, ferroelectric materials are not only prototypical systems for studying spontaneous symmetry breaking and structural phase transitions, but also key components for non-volatile memory devices, piezoelectric sensors, photocatalysis, and many other technologically important applications~\cite{Li2018,Scott1989,fang2018}. 
Driven by the need for further miniaturization of electronic devices, researchers have devoted significant efforts to reduce the thickness of thin-films ferroelectrics~\cite{setter2006,Park2015,Boscke2011}. 
Despite the depolarization field~\cite{Junquera2003,mehta1973,Wurfel1973}, which usually inhibits the electric polarization of thin-film ferroelectrics, a few groups have demonstrated ferroelectricity sustains in bulk ferroelectrics with thickness down to $\sim$ 1 nm\cite{Fong2004,Lee2019}. 
The recent discovery of ferroelectricity in monolayer or few-layer Van der Waals (vdW) materials offer new opportunities for shrinking the size of ferroelectric devices to the atomically thin regime.~\cite{Fei2018,Liu2016,Chang2016,Zhou2017,Sharma2019} 
Compared to conventional bulk ferroelectrics, a key advantage of two-dimension vdW materials is free of dangling bonds on the surface.
First-principles calculations also show that a large number of two-dimensional (2D) materials are piezoelectric.~\cite{Dong2017,blonsky2015,duerloo2012} 
Remarkably, some of them\cite{Fei2016,Dong2017} even demonstrate giant piezoelectric effects, which can be more than two-orders-of-magnitude stronger than bulk piezoelectric materials.

Currently, our understanding of the fundamental physical properties of 2D piezoelectric and ferroelectric systems is in an early stage, and the lack of a robust and economical fabrication process for high-quality 2D ferroelectric samples hinders mass production and applications~\cite{Cui2018}. 
Among 2D ferroelectric materials predicted with first-principles theories~\cite{Wu2016, Mehrshad2016, Fei2016,gao2019,Ding2017,anand2017,Lin201908}, only a few, such as monolayer SnS~\cite{Higashitarumizu2020}, SnSe~\cite{chang2020},  SnTe~\cite{Chang2016} and In$_2$Se$_3$~\cite{Zhou2017}, have so far been synthesized and confirmed to be ferroelectric. 
First-principles prediction of piezoelectricity or switchable electric polarization in readily fabricated 2D materials is important for enriching the toolbox of 2D non-centrosymmetric materials with technological interests.

Through first-principles calculations, we show ample evidence that three monolayer arsenic chalcogenides (As$_2$X$_3$) with the Pmn2$_1$ space group will exhibit spontaneous and reversible in-plane polarization. 
Among these three materials, the Pmn2$_1$ As$_2$S$_3$, i.e., monolayer orpiment, has recently been isolated through mechanical exfoliation~\cite{siskins2019}.
Moreover, we predict the existence of several novel metastable polymorphs of As$_2$S$_3$. 
Both ferroelectric monolayer orpiment and these new polymorphs can be related to the soft zone-center modes of a hypothetical high-symmetry phase. 
Remarkably, our calculations show some of these polymorphs have large piezoelectric coefficients comparable to those of group IV-VI compounds~\cite{Fei2016}. 

\section{Results and Discussion}

\begin{figure}[htb]
    \centering
    \includegraphics[width=0.45\textwidth]{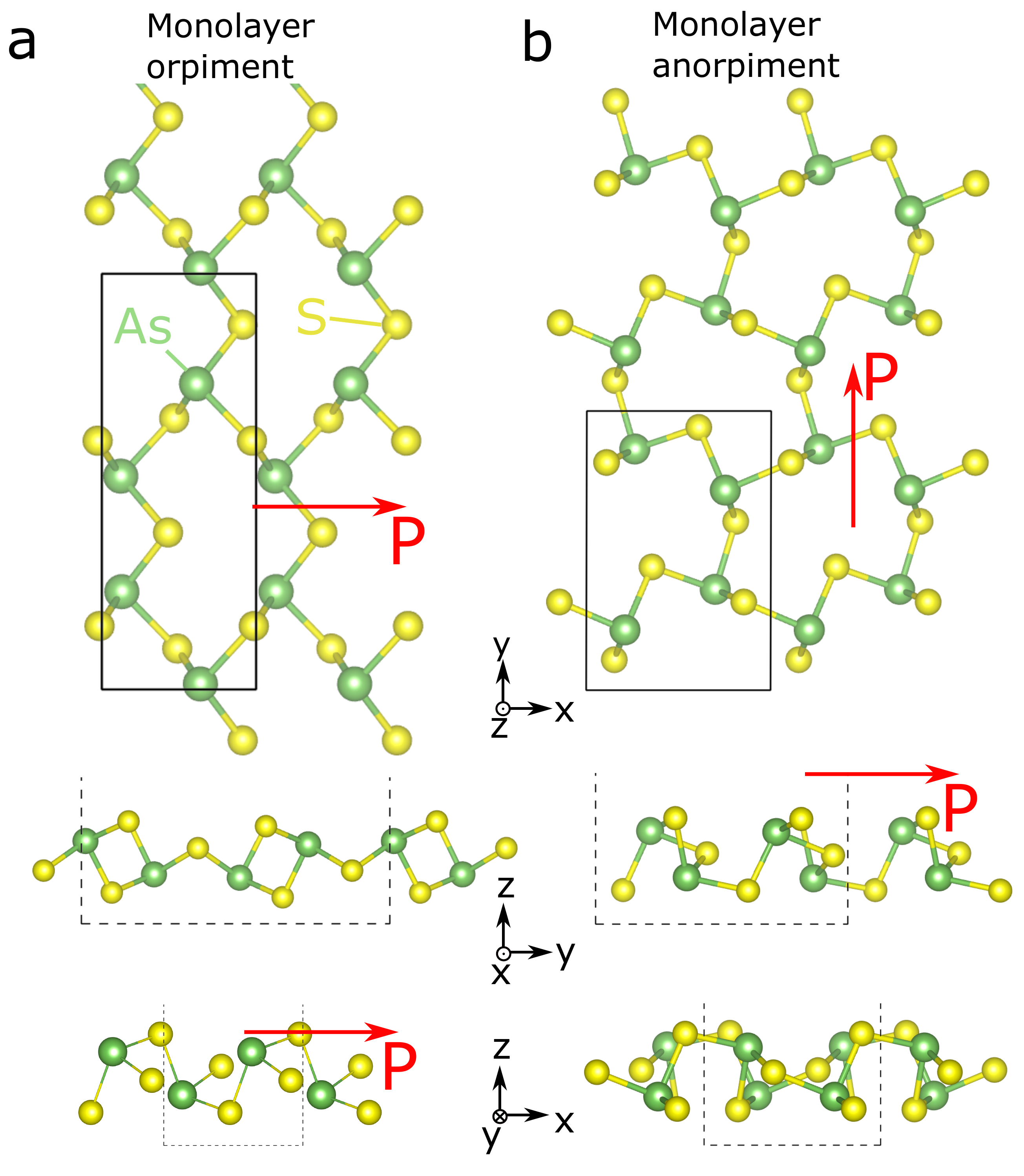}
    \caption{The crystal structures of (a) monolayer orpiment and (b) monolayer anorpiment, plot with VESTA\cite{Momma2011}. The red arrows show the direction of polarization.}
    \label{fig:1}
\end{figure}

\begin{figure*}[tb]
    \centering
    \includegraphics[width=0.75\textwidth]{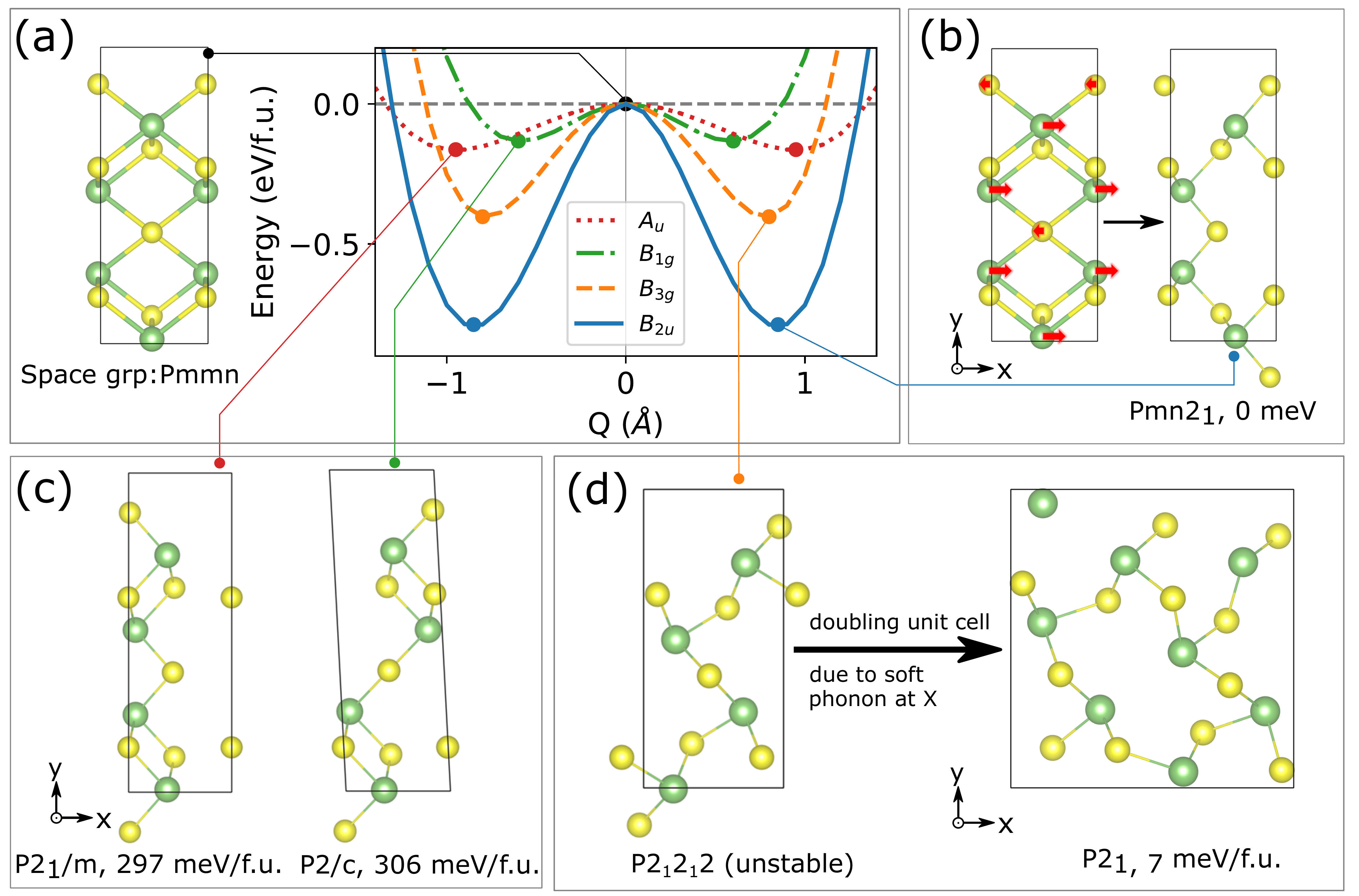}
    \caption{(a) (left) The high-symmetry structure with space group Pmmn and (right) the change of total energy under collective atomic displacements of the Pmmn structure with frozen-in soft phonon mode A$_u$, B$_{1g}$, B$_{3g}$, and B$_{2u}$. The displacement vector $\Delta \mathbf{R}_p$ is proportional to the polarization vector of phonon $p$: $\Delta \mathbf{R}_p = Q\cdot \mathbf{u}_p$. (b) Schematic plot of the transform from Pmmn structure to monolayer orpiment through the B$_{2u}$ soft mode. The red arrows (not to scale) show the moving direction of corresponding atoms. (c) Unit cell of the metastable P2$_1$/m phase and P2/c phase. (d) Schematic plot of the stabilization of the P2$_1$2$_1$2 phase by doubling the unit cell due to soft modes at $X$ point.}
    \label{fig:2}
\end{figure*}

Under ambient conditions, bulk As$_2$S$_3$ can be either amorphous or crystalline. 
Bulk orpiment and anorpiment, which were found in natural minerals~\cite{kampf2011}, are two common crystalline As$_2$S$_3$ phases with noncentrosymmetric layered structures bounded by vdW interactions. 
To date, two-dimensional anorpiment has not been synthesized, while monolayer and few-layer orpiment have been successfully exfoliated and demonstrates better chemical stability than phosphorene under low light conditions~\cite{siskins2019}. 
Our calculated total energy of monolayer orpiment is lower than that of monolayer anorpiment by 73~meV/formula unit (f.u.), suggesting better stability of monolayer orpiment compared to monolayer anorpiment. 
A finite bandgap is required for sustaining the ferroelectricity of 2D materials with in-plane polarization.
Monolayer orpiment and anorpiment have indirect bandgaps around 2.2 eV calculated with the Perdew-Burke-Ernzerhof (PBE) functional~\cite{PBE}. 
The band structures are presented in the Supporting Information.

The crystal structures of monolayer arsenic chalcogenides do not resemble those of other well-known 2D materials.
As shown in Fig.~\ref{fig:1}(a), monolayer orpiment is highly anisotropic and consists of rings connected by six corner-sharing AsS$_3$ units, which have a pyramidal shape. 
Monolayer orpiment has Pmn2$_1$ symmetry, which includes a mirror-reflection to the $xz$-plane, but no symmetry with the $yz$-plane, as illustrated in Fig.~\ref{fig:1}(a). 
Such symmetry properties allow a spontaneous electric polarization along the $x$-axis. 
In comparison, monolayer anorpiment has a more irregular structure and electric polarization pointing in the $y$-direction, as illustrated in Fig.~\ref{fig:1}(b).

Bulk As$_2$Se$_3$ can be found in mineral laphamite with a similar structure as orpiment~\cite{stergiou1985}, while bulk As$_2$Te$_3$ with the orpiment-like structure is yet to be found. 
Our calculations show monolayer As$_2$Se$_3$ and As$_2$Te$_3$ with the orpiment-like structure (Pmn2$_1$ symmetry) are dynamically stable, while those with the anorpiment-like structure (Pc space group) demonstrate dynamical instability with imaginary phonon modes. 

Using first-principles methods based on modern polarization theory~\cite{resta1994,king1993}, we calculate that monolayer orpiment has a spontaneous electric polarization of 71 pC/m. 
According to the experimental structure of bulk orpiment\cite{mullen1972,kampf2011}, the electric polarization in neighboring layers aligns in an antiferroelectric order. 
Therefore, bulk orpiment shows no macroscopic polarization and the net polarization of a few-layer orpiment shows an odd-even effect. Only samples with odd numbers of layers show net electric polarization. 

\begin{figure*}[htb]
    \centering
    \includegraphics[width=0.85\textwidth]{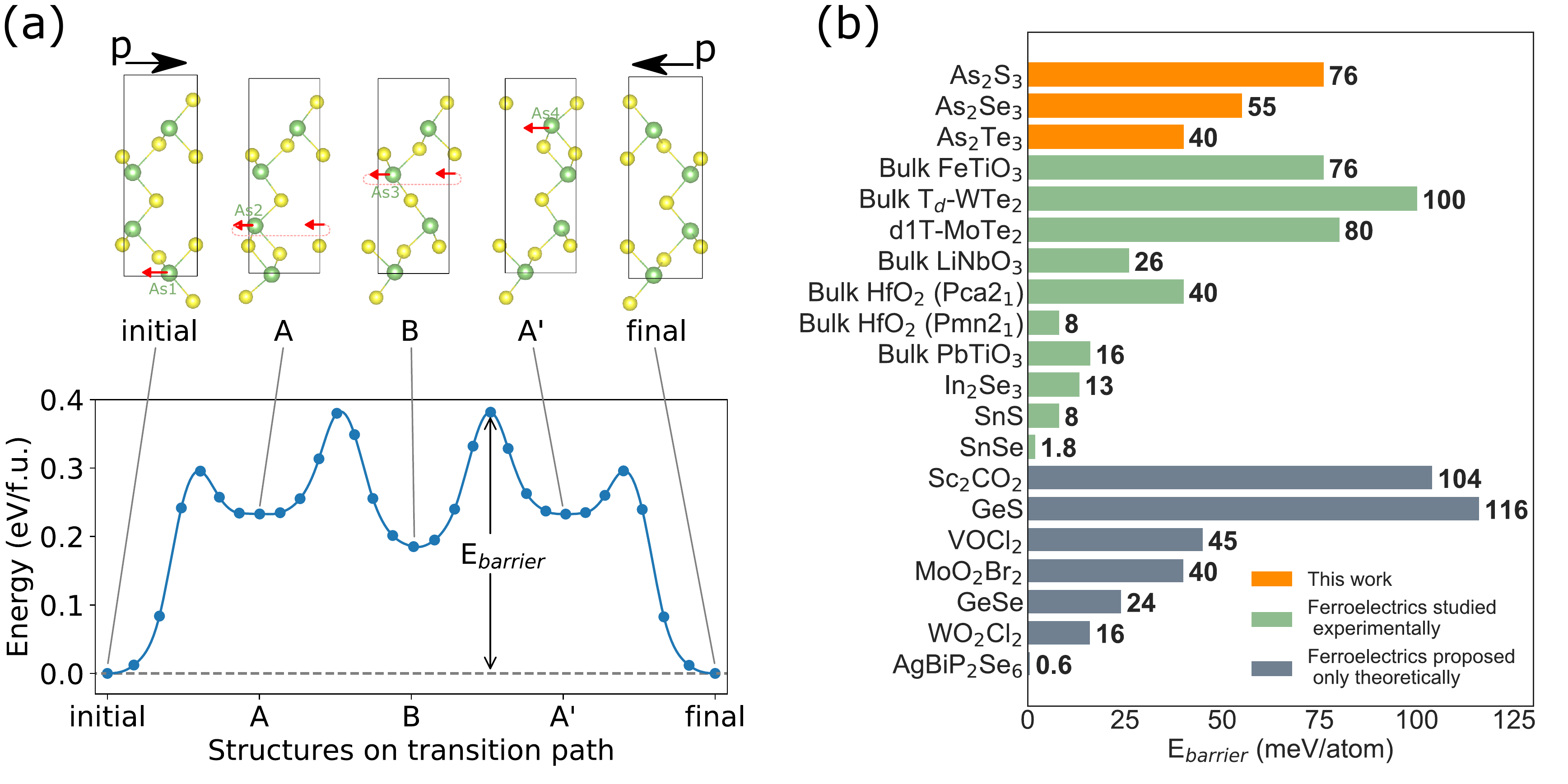}
    \caption{(a). (top) Selected intermediate states on the transition path of inverting the polarization direction of monolayer orpiment. Red arrows show the main movement of As ions between consecutive intermediate structures. The dashed curves are used to represent the atomic displacements that cross the unit-cell boundary. (bottom) The evolution of the total energies of intermediate structures along the transition path. (b). A comparison between the theoretical energy barriers $E_{barrier}$ of the polarization-reversing process of arsenic chalcogenide and other ferroelectrics. The values from previous work are all calculated with density functional theory.
    }
    \label{fig:3}
\end{figure*}

As a classical example of displacive transitions, the ferroelectric phase transition of perovskite oxide like PbTiO$_3$ is explained by a zone-center vibrational mode which vanishes at the phase transition.
Similarly, we propose the ferroelectricity of monolayer orpiment is also driven by a soft mode of a high-symmetry structure with space group Pmmn. 
The unit cell of the Pmmn structure is shown schematically in Fig.~\ref{fig:2}(a). 
Different from monolayer orpiment which only has mirror symmetry  to $xz$-plane, the Pmmn structure of As$_2$S$_3$ has additional mirror symmetry with $yz$-plane.
This high-symmetry Pmmn structure is dynamically unstable with five soft optical phonon modes at the $\Gamma$ point. 
To quantify contributions of a zone-center soft phonon mode to the structural transition from the high-symmetry Pmmn structure to Pmn2$_1$ phase, we calculate the projection of the atomic displacement vector $\Delta \mathbf{R} = \mathbf{R}_{Pmmn}-\mathbf{R}_{Pmn2_1}$ on soft phonon modes: 
\begin{equation*}
    \eta[p] = \frac{\Delta \mathbf{R}}{ |\Delta \mathbf{R}| } \cdot \mathbf{u}_p 
\end{equation*}
where $\mathbf{u}_p$ is the normalized polarization vector of a zone-center phonon $p$. 
We find $\eta[B_{2u}]=86\%$, and other four soft zone-center modes contribute less than 1 percent to $\Delta \mathbf{R}$. 
This is expected since the B$_{2u}$ mode is the only one that breaks the inversion symmetry and also has the deepest double-well potential curve among all zone-center soft modes. 
Therefore the B$_{2u}$ mode is the dominant phonon mode driving the structural transition from the unstable Pmmn structure to the Pmn2$_1$ phase of As$_2$S$_3$. 
As shown in Fig.~\ref{fig:2}(b), the main effect of B$_{2u}$ optical mode is to shift the As atoms along the $x$-axis and break the reflection symmetry to $yz$-plane.
The relative shifts between different S atoms are smaller comparing to the displacement of As atoms in the B$_{2u}$ mode. 

We also examine the roles of other four soft zone-center modes, namely B$_{3g}$, B$_{1g}$, B$_{3u}$, and A$_u$ modes, by collectively displacing atomic coordinates of the high-symmetry Pmmn structure by $\Delta \mathbf{R}_p = Q\cdot \mathbf{u}_p$, where $\mathbf{u}_p$ is the normalized polarization vector of phonon mode $p$.
As shown in Fig.~\ref{fig:2}(a), one can easily identify structures that correspond to the local minima (shown as small spheres) on the double-well curve of total energy versus general coordinate $Q$. 
Further relaxing local minimal structures may lead to new metastable phases of As$_2$S$_3$. 
Since the structural relaxation moves the local minima that correspond to the B$_{3u}$ mode back to monolayer orpiment, we will not discuss it further.

Interestingly, B$_{3g}$ and B$_{1g}$ modes transform the high-symmetry Pmmn structure into metastable P2/c and P2$_1$/m phases, respectively. 
As shown in Fig.~\ref{fig:2}(c), both P2/c and P2$_1$/m phase show unusual one-dimensional chain structures consist of interconnected AsS$_3$ pyramidal units. 
These two phases show zero macroscopic polarization since the dipole moments of neighboring AsS$_3$ pyramidal units point in opposite directions and thus cancel with each other.

A more complicated case is the soft A$_u$ mode.
Relaxing the local minima structure corresponding to the A$_u$ mode leads to a dynamically unstable P2$_1$2$_1$2 structure without a net electric polarization.
Such an unstable structure has doubly-degenerate soft phonon modes at the $X$ point which can stabilize the structure by doubling the unit cell along $x$-axis.
The final stable structure we find has the P2$_1$ space group symmetry and a rectangle unit cell with 20 atoms, as shown in Fig.~\ref{fig:2}(d).
The P2$_1$ phase has a noncentrosymmetric structure with spontaneous polarization of 20 pC/m pointing in the $y$-direction. 
We confirmed the stability of P2/c, P2$_1$/m, and P2$_1$ phases from their phonon spectra calculated with density functional perturbation theory~\cite{dfpt} and finite-temperature molecular dynamics~\cite{berendsen1984} trajectories. These results are presented in Supporting Information.
  
The total energy of the P2$_1$ phase As$_2$S$_3$ is 65 meV/f.u. lower than that of monolayer anorpiment, and only 7 meV/f.u. higher than that of monolayer orpiment. 
Even though the P2/c and P2$_1$/m phases are shown to be metastable, they have total energies which are about 300 meV/f.u. higher than that of monolayer orpiment, because they are composed with one-dimensional chain-like structures bounded by weak vdW forces. 
We mention only the zone-center soft modes of Pmmn structures are studied in this work. 
It is also interesting to study the finite-momentum soft modes, which also appear in the phonon spectrum of the Pmmn structure and may lead to other interesting polymorphs.

We perform similar analyses on the soft modes of the high-symmetry Pmmn structure of As$_2$Se$_3$ and As$_2$Te$_3$. 
Like As$_2$S$_3$, they both have a B$_{2u}$ mode driving the displacive transition to the corresponding Pmn2$_1$ phase. As$_2$Se$_3$ also has a metastable phase with P2$_1$ space group. 
In Table~\ref{tab:phase}, we list the space groups and the electric polarization of all stable polymorphs studied in this work. 
We find $P$ of Pmn2$_1$ phases decreases as the chalcogen element changes from sulfur to tellurium.
A similar trend also appears in IV-VI monolayers\cite{Fei2016}.
We explain this qualitatively with two arguments. 
First, The electrical polarization P is positively correlated to the difference between the electron negativity of As and the chalcogen elements. As the chalcogen element changes from S to Te, the reduced electronegativity results in a diminished polarization. Second, since the distortion amplitude $|Q_{min}|$ at the minima of the double-well potential of B$_{2u}$ mode decreases as the chalcogen element changes from S to Te, the dipole moment, which is proportional to $|Q_{min}|$, also decreases.

\begin{table}[htb]
\setlength\tabcolsep{0.02\textwidth}
\caption{Summary of the space group and electric polarization $P$ of different arsenic chalcogenides phases.}
\label{tab:phase}
\begin{tabular}{@{}ccc@{}}
\toprule
Formula & Space group & \begin{tabular}[c]{@{}c@{}}$P$ (pC/m) \\ (The direction of P)\end{tabular} \\ \midrule
\multirow{5}{*}{As$_2$S$_3$} & \begin{tabular}[c]{@{}c@{}}Pmn2$_1$\\ (Monolayer \\ orpiment)\end{tabular} & 71 (x) \\
 & \begin{tabular}[c]{@{}c@{}}Pc\\ (Monolayer \\ anorpiment)\end{tabular} & 47 (y) \\
 & P2/c & 0 \\
 & P2$_1$/m & 0 \\
 & P2$_1$ & 20 (y) \\ \midrule
\multirow{2}{*}{As$_2$Se$_3$} & Pmn2$_1$ & 54 (x) \\
 & P2$_1$ & 18 (y) \\ \midrule
As$_2$Te$_3$ & Pmn2$_1$ & 45 (x) \\ \bottomrule
\end{tabular}
\end{table}

The reversibility of electric polarization is a necessary condition for ferroelectrics and also important for the application in data storage. 
Using the nudged-elastic-band method~\cite{neb}, we find a minimal-energy-barrier transition path for reversing the electric polarization of an infinite large monolayer orpiment. 
We show important intermediate structures and the corresponding energies on the transition path of As$_2$S$_3$ in Fig.~\ref{fig:3} (a). 
In the polarization-reversing process, three important intermediate structures labeled as $A, B,$ and $A^\prime$ are found. 
$A$ and $A^\prime$ are related by a 180$^\circ$ rotation around the $z$-axis. 
$B$ is structurally akin to the unstable P2$_1$2$_1$2 structure. 
The initial and final structures on the transition path correspond to monolayer orpiment with electric polarization pointing in opposite directions.
The process of reversing electric polarization goes in the sequence  $\{\text{initial}\rightarrow A \rightarrow B \rightarrow A^\prime \rightarrow \text{final} \}$, which consists of four major steps. 
Each step mainly involves shifting a single As atom along the $x$-axis.
For example, in the first step $\{\text{initial}\rightarrow A\}$, the major structural change is the displacement of As1 (i.e. the arsenic atom at the bottom of the unit cell shown in Fig.~\ref{fig:3} (a)) along the negative $x$-direction.
Similarly, in the second step $\{A\rightarrow B\}$, we observe the movement of As2 along the negative $x$-direction to pass the boundary of the unit cell. 
We emphasize that the switching process we presented here may not correspond to the global minimum barrier, since the changes of lattice vectors are not considered in our nudged-elastic-band calculations and the electric dipoles can never be switched simultaneously in real situations.
Previous work also shows that including the variation of lattice vectors can further lower the energy barrier\cite{Salvador2018,Mehrshad2016}. 
Nevertheless, our theoretical energy barrier E$_{barrier}$ provides an upper bound for the activation energy of the real polarization-reversing process. 
With the similar approach, the energy profiles for reversing the electric polarization of As$_2$Se$_3$ and As$_2$Te$_3$ are calculated and shown in the Supporting Information. 

In Fig.~\ref{fig:3} (b), we compare calculated energy barriers E$_{barrier}$ of As$_2$X$_3$ with those of other ferroelectrics, which are either studied experimentally\cite{Sharma2019,Yuan2019,Huan2014,Sang2015,Higashitarumizu2020,Wang_2017,Cui2018_b,Zhou2017,Ding2017,Beckman2009,Ye2016,Brehm2020} or solely predicted in theory\cite{anand2017,Wang_2017,Lin201908,Ai2019,Xu2017,Jia2019}. 
Except FeTiO$_3$, LiNbO$_3$, AgBiP$_2$Se$_6$, and PbTiO$_3$, the E$_{barrier}$ in Fig.~\ref{fig:3} are calculated with the nudged-elastic-band method. 
The range of calculated $E_{barrier}$ covers two orders of magnitudes from 0.6 meV to 116 meV. 
The $E_{barrier}$ of As$_2$X$_3$ are notably smaller than those of two room-temperature ferroelectrics, namely $T_d$-WTe$_2$\cite{Sharma2019} and monolayer d1T-MoTe$_2$\cite{Yuan2019}, and a few predicted ferroelectrics such as GeS\cite{Wang_2017} and Sc$_2$CO$_2$\cite{Mehrshad2016}, but much larger than those of CuInP$_2$S$_6$\cite{Brehm2020}, In$_2$Se$_3$\cite{Ding2017}, SnS\cite{Wang_2017} and so on.
Such comparisons suggest that the energy barriers of switching the polarization direction of Pmn2$_1$ As$_2$X$_3$ are within a proper range.

\begin{figure}[htb]
    \centering
    \includegraphics[width=0.47\textwidth]{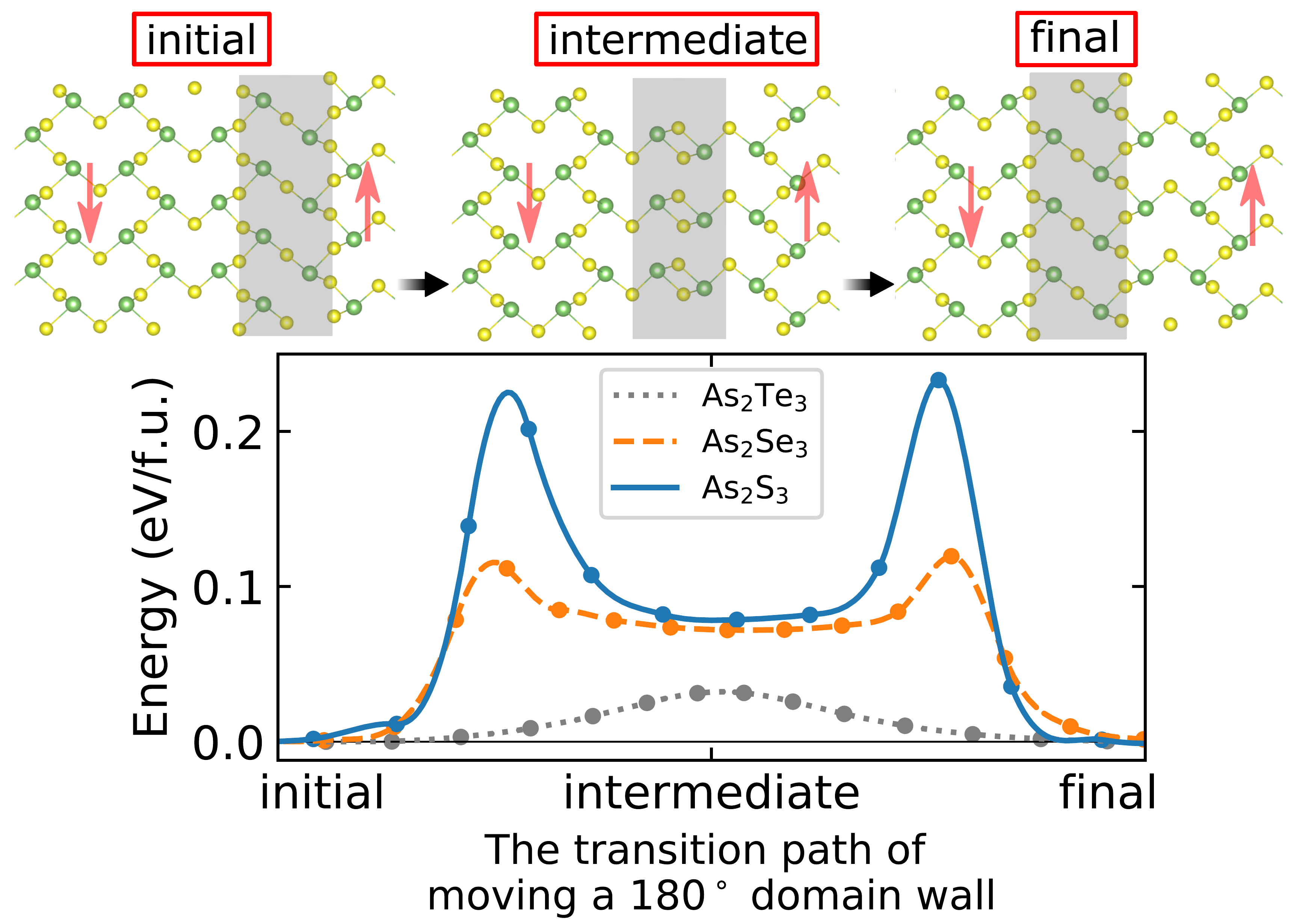}
    \caption{(top panel) Selected structures on the process of moving the 180$^\circ$ domain wall in As$_2$S$_3$. The shaded regions highlight the domain boundaries. (bottom panel) Total energies on the process of moving the domain walls of As$_2$S$_3$, As$_2$Se$_3$, and As$_2$Te$_3$.}
    \label{fig:4}
\end{figure}

In practical situations, domain-wall shifting and domain growing mediate the process of reversing electric polarization of ferroelectrics. 
Formation energies of domain walls and energy barriers for moving domain walls indicate how difficult it is to form and grow a domain, respectively. 
We studied the atomistic structure of a few 180$^\circ$ domain walls parallel to the $x$-axis in As$_2$X$_3$.
For example, the structure of the 180$^\circ$ domain wall in As$_2$S$_3$ is shown schematically in the top panel of Fig.~\ref{fig:4}. 
Our calculations indicate the energy costs of forming and moving the 180$^\circ$ domain wall along $x$-axis are reasonable compared to other ferroelectrics.
In detail, the calculated domain-wall formation energies $E^{dw}_{form}$ are 89, 105, and 124 meV/f.u. (i.e., 43, 50, and 58 meV/\AA) for monolayer As$_2$S$_3$, As$_2$Se$_3$, and As$_2$Te$_3$, respectively. 
Previous calculations show the $E^{dw}_{form}$ of group IV-VI materials range from 8 meV/$\text{\AA}$ to 116 meV/\AA~\cite{Wang_2017}, which covers those of 2D As$_2$X$_3$ ferroelectrics.
The $E^{dw}_{form}$ of In$_2$Se$_3$ is 220 meV/f.u.~\cite{Ding2017}, comparable to those of As$_2$X$_3$. 
Assuming the thickness of monolayer As$_2$X$_3$ is 6.0 \AA, we convert the $E^{dw}_{form}$ of As$_2$S$_3$, As$_2$Se$_3$, and As$_2$Te$_3$ to be 115, 133, and 155 mJ/m$^2$, which are in the same order as those of some ferroelectric oxides, such as PbTiO$_3$ (132 mJ/m$^2$ for 180$^\circ$ domain wall and 35.2 mJ/m$^2$ for 90$^\circ$ domain wall)~\cite{meyer2002} and BiFeO$_3$ (205 to 1811 mJ/m$^2$)~\cite{lubk2009}, but much higher than that of BaTiO$_3$ (7.5 mJ/m$^2$)~\cite{meyer2002}.

Using nudged-elastic-band method, we calculate the energy barriers $E^{dw}_{barrier}$ for moving the 180$^\circ$ domain walls are 233 meV/f.u., 128 meV/f.u., and 35 meV/f.u. (i.e., 113 meV/\AA, 54 meV/\AA, and 16 meV/\AA) for monolayer As$_2$S$_3$, As$_2$Se$_3$, and As$_2$Te$_3$, respectively, as shown in the bottom panel of Fig.~\ref{fig:4}. 
This suggests the 180$^\circ$ domain wall of As$_2$X$_3$ becomes easier to shift as the chalcogen element X changes from sulfur to tellurium. 
Compared to bulk ferroelectrics, the $E^{dw}_{barrier}$ of monolayer As$_2$Te$_3$ and As$_2$Se$_3$ are of the same order-of-magnitude as those of bulk corundum derivatives ranging from 14 meV/f.u. to 197 meV/f.u.~\cite{Meng2017}.
Compared to other two-dimensional ferroelectrics, $E^{dw}_{barrier}$ of monolayer As$_2$S$_3$ is more than an order-of-magnitude higher than those of group IV-VI two-dimensional ferroelectrics (less than 1.6 meV/\AA)~\cite{Wang_2017}, but in similar order with that of monolayer In$_2$Se$_3$, which ranges from 280 meV/f.u. to 400 meV/f.u.~\cite{Ding2017}. 
These comparisons suggest that the energy costs for forming and moving the 180$^\circ$ domain wall of As$_2$X$_3$ are reasonable.

\begin{table*}[htb]
\setlength\tabcolsep{0.018\textwidth}
\caption{Elasticity tensor elements (N/m) and piezoelectric coefficients ($10^{-10}$ C/m for $e_{ij}$ and pm/V for $d_{ij}$). }
\label{tab:piezo}
\begin{tabular}{@{}ccccccccc@{}}
\toprule
Space group (Point group) & Formula & $C_{11}$ & $C_{22}$ & $C_{12}$ & $e_{11}$ & $e_{12}$ & $d_{11}$ & $d_{12}$ \\ \midrule
\multirow{3}{*}{Pmn2$_1$ (C$_{2v}$)}  & As$_2$S$_3$ & 11.07 & 43.38 & 10.40 & 4.36 & -1.75 & 55.7 & -17.4 \\
 & As$_2$Se$_3$ & 13.86 & 41.76 & 10.03 & 6.71 & -1.49 & 61.7 & -18.4 \\
 & As$_2$Te$_3$ & 18.09 & 34.65 & 9.72 & 9.09 & -1.48 & 61.9 & -21.6 \\ \midrule
 &   & $C_{11}$ & $C_{22}$ & $C_{12}$ & $e_{21}$ & $e_{22}$ & $d_{21}$ & $d_{22}$ \\ \midrule
\multirow{2}{*}{P2$_1$ (C$_2$)}  & As$_2$S$_3$ & 18.63 & 16.29 & 3.42 & -0.85 & 0.22 & -5.0 & 2.4 \\
 & As$_2$Se$_3$ & 21.51 & 23.76 & 2.65 & -0.91 & -0.33 & -4.1 & -0.9 \\ \midrule
Pc (C$_s$) & As$_2$S$_3$ & 21.63 & 9.25 & 8.00 & -2.77 & 2.69 & -34.7 & 59.1 \\ \bottomrule
\end{tabular}
\end{table*}

Similar to monolayer group IV-VI compounds and black phosphorene, monolayer arsenic chalcogenides studied in this work are super flexible. 
We calculate Young's modulus of monolayer orpiment to be 8.6~N/m along the $x$-axis and 33.6~N/m along the $y$-axis, which is more than one-order-of-magnitude smaller than those of graphene (345~N/m)\cite{kudin2001,deji2017} and also significantly smaller than that of black phosphorene ($21\sim56$ N/m)\cite{Jiang2014_b,wei2014}.
To our knowledge, monolayer orpiment is among the softest 2D material ever fabricated.
Such remarkable structural flexibility motivates us to investigate the piezoelectricity of arsenic chalcogenides. 

We summarize the calculated elasticity tensor $C_{ij}$ and piezoelectric tensor elements $e_{ij}$ and $d_{ij}$  in Table~\ref{tab:piezo}. 
More details of calculating these tensor elements are presented in Supporting Information.
Obviously, the piezoelectric strain coefficients $d_{ij}$ of Pmn2$_1$ and Pc phases are one-order-of-magnitude larger than those of common two-dimensional polar materials such as 2H-MoSe$_2$ ($d_{11}=3.73$ pm/V)~\cite{duerloo2012}, 2H-WSe$_2$ ($d_{11}=2.79$ pm/V)~\cite{duerloo2012}, hexagonal group III-V materials ($0.02<d_{11}<5.50$ pm/V)~\cite{blonsky2015}, and multilayer janus transition metal chalcogenide MoSTe ($5.7<d_{33}<13.5$ pm/V)~\cite{Dong2017}. 
On the other hand, the piezoelectric stress constants $e_{ij}$ of arsenic chalcogenides are comparable with those of 2H-MoSe$_2$, 2H-WSe$_2$, and so on~\cite{duerloo2012,blonsky2015,Dong2017}.
This indicates the large $d_{ij}$ coefficients of arsenic chalcogenides is originated from their superior flexibility, i.e., small elasticity tensor components.
Compared to group IV-VI monolayers with giant piezoelectricity, the $d_{ij}$ coefficients of Pmn2$_1$ and Pc phases are on the same order as that of GeS, but two to four-times smaller than those of SnS, SnSe, and GeSe~\cite{Fei2016}. 
The piezoelectric coefficients of P2$_1$ are much smaller than other phases. Interestingly, P2$_1$ As$_2$Se$_3$ shows weak negative piezoelectric effect along y-direction.

In summary, we employ ab initio methods to predict the intrinsic ferroelectricity and strong piezoelectricity in arsenic chalcogenides, which include the recently isolated monolayer orpiment. 
By analyzing the soft optical modes of the high-symmetry Pmmn structures of arsenic chalcogenides, we find these soft modes can lead to several undiscovered metastable polymorphs. 
The Pmn2$_1$ ferroelectric phases can be related to the soft B$_{2u}$ phonon mode of a high-symmetry Pmmn structure. 
We investigate the feasibility of switching the electrical polarization in the Pmn2$_1$ phase. 
The energy barrier of coherently flip all electrical dipoles and that of moving a 180$^\circ$-domain wall in two-dimensional Pmn2$_1$ As$_2$X$_3$ are in a proper range compared with other ferroelectrics.
Moreover, superior structural flexibility results in large piezoelectric responses in a few polymorphs. 
Such a unique combination of unusual structures, pliability, strong piezoelectricity, and predicted ferroelectricity make monolayer arsenic chalcogenides new platforms of studying polar materials. 
They are also convincing candidates for small-sized, flexible electronic devices.

Computation Details: Our first-principles calculations are based on pseudopotential density functional theory implemented in Quantum Espresso~\cite{QE-2017,QE-2009} and PARSEC~\cite{Chelikowsky1994,Kronik2006}. More technical details are presented in Supporting Information, which cites these references~\cite{PBEsol,Dalcorso2014,ultrasoft,troullier1991,dfpt,berendsen1984,mullen1972,Nye2012}. 

\acknowledgement
W.G. and J.R.C. acknowledge support from a subaward
from the Center for Computational Study of Excited-State
Phenomena in Energy Materials at the Lawrence Berkeley
National Laboratory, which is funded by the U.S. Department
of Energy, Office of Science, Basic Energy Sciences, Materials
Sciences and Engineering Division under Contract No. DEAC02-05CH11231,
as part of the Computational Materials
Sciences Program. 
Computational resources are provided by the Texas Advanced Computing Center (TACC).

\suppinfo
The supporting information presents more details on piezoelectric tensors, phonon spectra, molecular dynamics simulation, transition path for reversing electric polarization, and structure parameters of arsenic chalcogenides polymorphs. This material is available free of charge via the internet at http://pubs.acs.org.



\providecommand{\latin}[1]{#1}
\makeatletter
\providecommand{\doi}
  {\begingroup\let\do\@makeother\dospecials
  \catcode`\{=1 \catcode`\}=2 \doi@aux}
\providecommand{\doi@aux}[1]{\endgroup\texttt{#1}}
\makeatother
\providecommand*\mcitethebibliography{\thebibliography}
\csname @ifundefined\endcsname{endmcitethebibliography}
  {\let\endmcitethebibliography\endthebibliography}{}

\end{document}